\pdfoutput=1
\documentclass[12pt]{article}

\usepackage{graphicx}
\usepackage{subcaption}
\usepackage{colordvi}
\usepackage{fancybox}
\usepackage{cancel}
\usepackage{amsmath}
\usepackage{amsfonts}
\usepackage{slashed}
\usepackage{amssymb}
\usepackage{caption}
\usepackage{appendix}
\usepackage{hyperref} 
\usepackage{multirow}
\hypersetup{colorlinks=true, linkcolor=black, urlcolor=blue}
\usepackage{float}
\usepackage{mathtools}
\usepackage{bbm}
\usepackage{dsfont}

\usepackage[backend=biber, style=numeric-comp, sorting=none,
            doi=false, maxbibnames=99]{biblatex}
\newcommand{\beq}{\begin{equation}}
\newcommand{\eeq}{\end{equation}}
\newcommand{\bea}{\begin{eqnarray}}
\newcommand{\eea}{\end{eqnarray}}
\newcommand{\bem}{\begin{pmatrix}}
\newcommand{\eem}{\end{pmatrix}}

\newcommand{\bet}{\begin{itemize}}
\newcommand{\eet}{\end{itemize}}
\newcommand{\ben}{\begin{enumerate}}
\newcommand{\een}{\end{enumerate}}

\begin{document}

\newcommand{\ba}[1]{\begin{array}{#1}} \newcommand{\ea}{\end{array}}

\numberwithin{equation}{section}

\bigskip

\begin{center}

{\Large\bf Holographic thermal propagator at finite energy from modularity and triality
 }
\vspace{1cm}

\centerline{
Borut Bajc$^{a}$, Mendrit Latifi$^{b}$,
Katarina Trailovi\'c$^{a,b}$\footnote{katarina.trailovic@ijs.si}
}
\vspace{0.5cm}
\centerline{
$^{a}$ {\it\small J.\ Stefan Institute, 1000 Ljubljana, Slovenia}
}
\centerline{
$^{b}$ {\it\small Faculty of Mathematics and Physics, University of Ljubljana,}
}
\centerline{{\it\small 1000 Ljubljana, Slovenia}}
\end{center}

\bigskip

\begin{abstract}
We extend the modularity-based approach of \cite{Bajc:2025jjv} for the holographic thermal propagator to the physically relevant case of nonzero energy $\omega$. Via the AGT correspondence, the problem is mapped to $\mathcal{N}=2$ $SU(2)$ gauge theory with $N_f=4$ hypermultiplets. Turning on $\omega$ activates all four hypermultiplet masses, so that the full $SO(8)$ flavor symmetry constrains the low-temperature expansion, rather than the subgroup left unbroken at $\omega=0$. As a result, the building blocks of the expansion are no longer just Eisenstein series but also Jacobi theta functions. We use this fact, together with the Molien series of the resulting invariant ring and the modular anomaly equation, to fix the low-temperature expansion coefficients order by order, matching only a finite number of terms in the Zamolodchikov's $q$-recursion at each order. We obtain the low-temperature expansion of the propagator and find agreement with the result of a completely different method, the near-boundary expansion of the bulk wave equation.
\end{abstract}

\clearpage

\tableofcontents

\section{Introduction}
 
Thermal two-point functions of holographic CFTs encode a wealth of real-time physics of the dual black hole: they contain the multi-stress-tensor data of the thermal OPE \cite{Fitzpatrick:2019zqz}, and their singularities carry information about the black hole interior \cite{Fidkowski:2003nf,Festuccia:2005pi,Ceplak:2024bja}. Computing them at small but non-zero temperature has been approached in several complementary ways. These include the near-boundary expansion of the bulk wave equation \cite{Fitzpatrick:2019zqz}, supplemented by the KMS condition in \cite{Buric:2025anb,Buric:2025fye}, the ambient space formalism \cite{Parisini:2022wkb,Parisini:2023nbd}, direct analytic solutions of the bulk wave equation \cite{Policastro:2001yb,Bajc:2022wws}, numerical approaches \cite{Niarchos:2026wsw,Niarchos:2026vow}, and the exact solution \cite{Bonelli:2022ten} of the connection problem of the Heun equation \cite{Dodelson:2022yvn,Jia:2024zes,Jia:2026ryl}.

The last approach is fully explicit. Via the AGT correspondence \cite{Alday:2009aq}, the retarded propagator of the black brane is expressed through the Nekrasov-Shatashvili (NS) limit \cite{Nekrasov:2009rc} of the Nekrasov partition function \cite{Nekrasov:2002qd} of $\mathcal{N}=2$ $SU(2)$ gauge theory with $N_f=4$ fundamental hypermultiplets \cite{Seiberg:1994aj}. The parameters of the Heun equation become the hypermultiplet masses, the instanton counting parameter $q$, and the Coulomb branch parameter $u$, related to the adjoint vev $a$ by the Matone relation \cite{Matone:1995rx}. The resulting expression involves two infinite sums: one over powers of the temperature and one over instanton numbers. For the black brane the instanton counting parameter is fixed at $q=e^{-\pi}$, so none of the instanton contributions can be neglected. In \cite{Bajc:2025jjv} two of us showed how to organize this double expansion in the special case $\omega=0$: at fixed order in the temperature, all instanton contributions can be resummed into a quasi-modular form built from the Eisenstein series $E_2,E_4,E_6$. This relies on the known quasi-modularity of the $N_f=4$ theory and on the associated modular anomaly equation \cite{Billo:2013fi,Billo:2013jba}.\footnote{See \cite{Beccaria:2026ndg} for a similar use of quasi-modularity in the low-temperature expansion of the double-scaled SYK partition function.} Once the coefficients multiplying positive powers of $E_2$ are fixed by the anomaly equation, only finitely many coefficients remain at each order, and these are determined by matching to a finite number of terms of Zamolodchikov's $q$-recursion \cite{Zamolodchikov:1984eqp,Zamolodchikov:1987avt,Zamolodchikov:1995aa,Poghossian:2009mk}.
 
The purpose of this paper is to extend this method to the physically relevant case of non-zero energy, $\omega\neq0$. Two new features arise. The first is that at $\omega=0$ only two of the four hypermultiplet masses are non-zero, whereas for $\omega\neq0$ all four are. Consequently, odd powers of $q$ appear in the instanton expansion, and $E_2,E_4,E_6$ alone no longer suffice: the Jacobi theta functions $\theta_2^4,\theta_4^4$ also enter. Neither they nor the quartic $SO(8)$ mass invariants $T_1,T_2$ are separately modular covariant. We show, however, that the two pairs transform in the same way under the modular group, up to an automorphy factor, through a common permutation group $S_3$. On the mass side this $S_3$ is the triality outer automorphism of the $SO(8)$ flavour symmetry. On the Heun side it is the permutation of the singular points $0,t,1,\infty$, i.e.\ the anharmonic group acting on the cross-ratio $t=\theta_2^4/\theta_3^4$. Every $S_3$-invariant polynomial in $(T_1,T_2)$ and $(\theta_2^4,\theta_4^4)$ is therefore automatically a modular form. The Molien series \cite{Stanley:1979} of the invariant ring counts these invariants at each weight. Besides the known generators, it yields three genuinely new mixed invariants built from both the masses and the theta functions. This fixes the most general ansatz for the expansion coefficients, which is then determined order by order, as for $\omega=0$, by the modular anomaly equation and a finite number of terms of Zamolodchikov's recursion.
 
The second new feature is kinematic. In the low-temperature limit both $\omega/T$ and $|\vec k|/T$ become large; we keep their ratio $\omega/|\vec k|<1$ fixed. The large-$\omega$ behaviour, however, only matters when the Matone relation is inverted. The dependence on $\omega$ enters in two ways: explicitly, through the masses in the coefficients of the large-$a$ expansion, and through the Matone relation, which expresses $a^2$ in terms of both $k^2=\vec k^2-\omega^2$ and $\omega^2$. Only after this inversion does the result become a series in $T$, and since $\omega/T$ is as large as $|\vec k|/T$, a term carrying $\omega^{2m}$ contributes at a lower order in $T$ than its power of $1/a$ alone would suggest. For example, a term $\omega^2/a^6$ scales as $T^4\,\omega^2/k^6$, so it contributes at order $T^4$ rather than $T^6$.
 
With these ingredients we obtain the low-temperature expansion of the retarded propagator $G_R(\omega,\vec k)$ through order $T^{12}$.
The result agrees with the low-temperature expansion obtained from the near-boundary expansion of the bulk wave equation \cite{Fitzpatrick:2019zqz}.
This agreement tests the identification of the finite-$\omega$ black brane propagator with the $S$-duality structure of $\mathcal{N}=2$ $SU(2)$ SYM with $N_f=4$, now with all four masses switched on.
 
The paper is organized as follows. In Section~\ref{sec:prop} we review the map between the black brane propagator and the NS free energy and describe how the low-temperature expansion is obtained for $\omega\neq0$. In Section~\ref{sec:modular} we analyse how the mass invariants and theta functions transform under the modular group, construct the $S_3$ invariants using the Molien series, and write the general ansatz for the expansion coefficients. In Section~\ref{sec:zam} we fix the remaining coefficients by matching to Zamolodchikov's $q$-recursion. We present the final result and our conclusions in Section~\ref{sec:conclusions}. The appendices collect properties of Eisenstein series and theta functions (Appendix~\ref{app:EisensteinJacobi}), details of Zamolodchikov's recursion (Appendix~\ref{ap:Zamolodchikov}), a review of Molien series (Appendix~\ref{app:molien}), and higher-order coefficients (Appendix~\ref{app:alphatilde}).

\section{The retarded Green's function of the black brane \label{sec:prop}}
In \cite{Bajc:2025jjv} the low-temperature expansion of the holographic thermal propagator was obtained in the simplified kinematic regime $\omega=0$, where the propagator depends only on the momentum square $\vec{k}^2$. Here we turn on the energy $\omega$ and repeat the construction of \cite{Bajc:2025jjv} for the general case. As reviewed there, the retarded two-point function holographically dual to the black brane AdS is encoded in the connection problem of a Heun equation with singular points $z=0,t,1,\infty$, their exponents $(a_0,a_t,a_1,a_\infty)$ and the accessory parameter $u$, which for the black brane background in Lorentzian signature take the values
\begin{equation}
    t=\frac{1}{2}, \quad \vec{a}\equiv (a_0, a_t, a_1, a_\infty)= \left(0, \frac{i \omega}{4\pi}, \frac{\nu}{2},\frac{\omega}{4\pi}\right), \quad u=\frac{\omega^2-2\vec{k}^2}{8\pi^2}-\frac{\nu^2}{4},
\end{equation}
where $\nu=\Delta-2$ with $\Delta$ being the conformal dimension of the scalar operator dual to the bulk field. Here and in the rest of the paper, $\omega$ and $\vec{k}$ denote the dimensionless combinations $\omega/T$ and $\vec{k}/T$, i.e.\ energy and momentum are measured in units of the temperature $T$ of the black brane. The explicit dependence on $T$ is restored only at the very end, when we expand the propagator in powers of $T$.
Compared to \cite{Bajc:2025jjv}, two things change: $u$ now depends also on $\omega$, and $a_t,a_\infty$ are no longer zero. Correspondingly, two of the four hypermultiplet masses, which vanished at $\omega=0$, become nonzero once $\omega\neq0$, and it is precisely these masses that will be responsible for all the new structures appearing in the next section.

As shown in \cite{Bonelli:2022ten}, the connection problem of the Heun equation can be solved by identifying the local solutions of the differential equation around different singular points with the conformal blocks in different OPE channels of the Liouville CFT in the semiclassical limit. Hence, the crossing symmetry of the Liouville conformal blocks gives the connection coefficients for the Heun equation. Finally, using the AGT correspondence \cite{Alday:2009aq}, the Liouville conformal block is related to the instanton part of the Nekrasov partition function and the semiclassical limit of the Liouville theory translates into the Nekrasov-Shatashvili (NS) limit of the $\Omega$-background \cite{Nekrasov:2009rc}. As a result, the parameters of the Heun equation acquire a direct meaning in the NS limit of the $\mathcal{N}=2 \ SU(2)$ SYM gauge theory with $N_f=4$ hypermultiplets: fixed linear combinations of $(a_0,a_t,a_1,a_\infty)$ are the hypermultiplet masses, $t$ is related to the instanton counting parameter $q$ and $u$ labels to the moduli space of vacua, i.e.\ $u=\langle \text{Tr} \phi^2\rangle$, which is related to the vev of the scalar in the vector multiplet $\langle \phi\rangle=\text{diag}(a,-a)$ via the Matone relation \cite{Matone:1995rx} 
\begin{equation}
    u=-a^2+a_t^2+a_0^2-\frac{1}{4}+ t\partial_tF, 
\end{equation}
where $F=\lim_{(\epsilon_1, \epsilon_2)\rightarrow (1,0)} \epsilon_1 \epsilon_2 \log Z_{\text{inst}}$ is the instanton part of the NS free energy.
As we showed in \cite{Bajc:2025jjv}, the NS free energy can be written as
\begin{equation}
\begin{split}
    F(q,a,\vec{a})&=\left(a_1^2+a_t^2-\frac{1}{4}\right)\log(1-t(q))+a^2\log{\left( \frac{t(q)}{16q}\right)}\\ &+\left(a_t^2+a_0^2+a_1^2+a_\infty^2-\frac{1}{4}\right)\log{\theta_3^4(q)}+H(q,a,\vec{a}),
       \end{split}
\end{equation}
where 
\beq
t(q)=\frac{\theta_2^4(q)}{\theta_3^4(q)}
\eeq
is the ratio of Jacobi theta functions and 
\begin{equation}
H(q,a,\vec{a})=\lim_{\substack{\epsilon_1\rightarrow 1\\ \epsilon_2\rightarrow 0}}\epsilon_1 \epsilon_2 \log{\mathcal{H}(q,a,\vec{a})}
\label{HlimH}
\end{equation}
with $\mathcal{H}(q,a,\vec{a})$ being essentially the 4-point conformal block of the 2d CFT, stripped of a number of kinematic prefactors (see Appendix \ref{ap:Zamolodchikov} for details).

We are interested in the low-temperature expansion of the propagator, i.e.\ the limit $T\rightarrow 0$ at fixed ratio $\omega/|\vec{k}|<1$. In terms of the dimensionless variables introduced above, this means $\omega\rightarrow\infty$ and $|\vec{k}|\rightarrow\infty$ with $\omega/|\vec{k}|<1$ fixed. The Matone relation then implies $a\rightarrow\infty$, and more precisely $a\sim|\vec{k}|$.

Following \cite{Dodelson:2022yvn}, in this limit the propagator can be written, up to exponentially suppressed corrections, as
\begin{equation}
    G_R=\pi^{4 a_1} e^{-\partial_{a_1} F} \frac{M(a,a_1;a_\infty)}{M(a,-a_1;a_\infty)},
\end{equation}
where
\begin{equation}
M(\alpha_0,\alpha_1;\alpha_2)=\frac{\Gamma(-2\alpha_1)\Gamma(1-2\alpha_0)}{\Gamma(1/2-\alpha_0-\alpha_1+\alpha_2)\Gamma(1/2-\alpha_0-\alpha_1-\alpha_2)}.
\end{equation}
Before expanding, it is convenient to introduce the rescaled momentum
\begin{equation}
    z\equiv\frac{k}{\sqrt{8 \pi \Gamma(3/4)^4}}, \qquad k^2=\vec{k}^2-\omega^2 ,
    \label{paramz}
\end{equation}
where the expression for $k^2$ holds in Lorentzian signature. The normalization is chosen so that, by the Matone relation \eqref{LorMatone} below, $a\simeq z$ at leading order for large momentum. Expanding $G_R$ via the large-$a$ asymptotics of the Gamma functions, exactly as in \cite{Bajc:2025jjv}, we then obtain
\begin{equation}
    \frac{G_R}{G_R^0}=\left(\frac{a^2}{z^2}\right)^\nu \exp\left(-2 \sum_{n=1}^\infty \frac{\partial_\nu \tilde{\alpha}_{2n}(\nu, \omega)}{a^{2n}}\right),
    \label{propagator}
\end{equation}
where \footnote{As explained later in Section \ref{sec:zam}, $\tilde \alpha_{2n}=\alpha_{2n}+f_{2n}$ is the sum of the part $\alpha_{2n}$ coming from the instanton prepotential and the part $f_{2n}$ coming from the perturbative prepotential. }
\begin{equation}
    \partial_\nu \tilde \alpha_{2n}(\nu,\omega)\equiv\partial_\nu \alpha_{2n}(\nu,\omega)+\frac{B_{2n+1}(\frac{\nu+1}{2}-\frac{\omega}{4\pi})+B_{2n+1}(\frac{\nu+1}{2}+\frac{\omega}{4\pi})}{2n (2n+1)}
\end{equation}
with $B_{m}(x)$ the Bernoulli polynomials, and where the large-$a$ expansion of \eqref{HlimH} defines $\alpha_{2n}(\nu, \omega)$ through 
\begin{equation}
    H(a,\omega,\nu)=\sum_{n=1}^{\infty} \frac{\alpha_{2n}(\nu, \omega)}{a^{2n}}.
\end{equation}
Here $G_R^0$ is the conformal (zero-temperature) propagator used to normalize $G_R$ above,
\begin{equation}
    G_R^0=\frac{ \Gamma(-\nu)}{\Gamma(\nu)}\left(\frac{k}{2}\right)^{2\nu}.
\end{equation}
 The Matone relation in Lorentzian signature rewritten  in terms of $z$ is
 \begin{equation}
     z^2=a^2-\frac{\nu^2-1}{4\pi}-\frac{ \omega^2}{16 \pi \Gamma(3/4)^4}-\sum_{n=1}^\infty \frac{q \partial_q \tilde \alpha_{2n}}{a^{2n}},
     \label{LorMatone}
\end{equation}
while in Euclidean signature we have just to replace everywhere $\omega^2\to-\omega^2$. 

To turn \eqref{propagator} into an expansion in $z$ (and eventually in $T$), we invert \eqref{LorMatone},
\begin{equation}
    a^2=z^2+\sum_{n=0}^\infty \frac{\beta_{2n}}{z^{2n}},
    \label{inversMatone}
\end{equation}
by plugging the ansatz \eqref{inversMatone} back into \eqref{LorMatone} and matching order by order in $1/z^{2n}$, which fixes the coefficients $\beta_{2n}(\nu, \omega)$.
Once these coefficients together with $\tilde \alpha_{2n}(\omega, \nu)$ itself are known, the procedure to obtain the low-temperature
expansion of the propagator is fixed: 
insert \eqref{inversMatone} into \eqref{propagator} and expand for large $z$, then use \eqref{paramz} together with $k\rightarrow k/T$ and $\omega\rightarrow \omega/T$  to restore the temperature dependence, and finally expand in positive powers of $T$.

So the only missing ingredient, is the explicit form of $\tilde \alpha_{2n}(\omega,\nu)$. Its computation, however, is genuinely more involved than in \cite{Bajc:2025jjv}: for $\omega\neq0$ the relevant $q$-series is no longer purely a function of the Eisenstein series $E_{2,4,6}$, but also involves Jacobi theta functions. We turn to this in the next two sections: Section~\ref{sec:modular} constructs the general modular-covariant ansatz for $\tilde\alpha_{2n}(\omega,\nu)$, and Section~\ref{sec:zam} fixes the finitely many free coefficients in that ansatz by matching to the explicit $q$-expansion generated by Zamolodchikov's recursion.

\section{Modular invariants \label{sec:modular}}
In \cite{Bajc:2025jjv} only two of the four hypermultiplet masses were nonzero, and as a consequence the instanton coefficients $R_{m,n}$ of the Zamolodchikov recursion (reviewed in Appendix~\ref{ap:Zamolodchikov}) vanish whenever $m$ and $n$ are both odd. Due to this only even powers of $q$ appear, which is why the resulting $q$-series organizes itself purely in terms of the Eisenstein series $E_2,E_4,E_6$. Once $\omega\neq0$, however, all four masses are generically nonzero,
\begin{equation}
\label{mass}
    (m_1,m_2,m_3,m_4)=(a_t+a_0,a_t-a_0,a_1+a_\infty,a_1-a_\infty)=\Big(\frac{i\omega}{4\pi},\frac{i\omega}{4\pi},\frac{\nu}{2}+\frac{\omega}{4\pi},\frac{\nu}{2}-\frac{\omega}{4\pi}\Big)\, ,
\end{equation}
$R_{m,n}$ no longer vanishes for $m,n$ both odd, and odd powers of $q$ appear in the $q$-expansion of $H$. Correspondingly, the building blocks needed to write $\tilde\alpha_{2n}$ in closed form are no longer just $E_2,E_4,E_6$, but also the Jacobi theta functions $\theta_2,\theta_4$, together with the mass invariants that control how they combine. The purpose of this section is to construct the most general expression for $\tilde\alpha_{2n}(q)$ compatible with the two symmetries of the problem: invariance under the Weyl group of the $SO(8)$ flavour symmetry acting on the masses \eqref{mass}, and (quasi-)modular covariance under the modular group $PSL(2,\mathbb{Z})$ acting on the gauge coupling, and hence on $q$. The finitely many free coefficients in this expression will then be fixed in Section~\ref{sec:zam} by matching to the explicit low-order $q$-expansion.

The $\mathcal{N}=2$ $SU(2)$ SYM with $N_f=4$ fundamental hypermultiplets has an $SO(8)$ flavor symmetry, whose Weyl group acts on the masses by permutations together with an even number of sign flips. Physical quantities such as the prepotential must therefore be invariant under this action of $SO(8)$ \cite{Seiberg:1994aj}. The quadratic, quartic and sextic $SO(8)$ invariants are \cite{Billo:2013fi}: \footnote{Since the Weyl group of $SO(8)$ acts as a reflection group on the four masses, its invariant ring is freely generated by exactly four invariants, of degrees $2,4,4,6$ -- the Coxeter degrees of $D_4$. The double degree-$4$ entry, unique to $D_4$, is why there are two independent quartic invariants $T_1,T_2$: it is a signature of triality.}
\begin{align}
\label{R}
R&=\frac{1}{2} \sum_f m_f^2=\frac{\nu^2}{4},\\
\label{T1}
T_1&=\frac{1}{12} \sum_{f<f'} m_f^2 m_{f'}^2-\frac{1}{24} \sum_f m_f^4=-\frac{12 \pi^2 \nu^2\omega^2+\omega^4}{768 \pi^4},\\
\label{T2}
T_2&=-\frac{1}{24} \sum_{f<f'} m_f^2 m_{f'}^2+\frac{1}{48} \sum_f m_f^4-\frac{1}{2} \prod_f m_f=-\frac{-12 \pi^2 \nu^2\omega^2+\omega^4}{768 \pi^4}, \\
\label{N}
N&=\frac{3}{16} \sum_{f<f'<f''} m_f^2 m_{f'}^2 m_{f''}^2-\frac{1}{96}\sum_{f\neq f'} m_f^2 m_{f'}^4+\frac{1}{96} \sum_f m_f^6=\frac{5 \nu^2 \omega^4}{3072 \pi^4}.
\end{align}
As discussed in \cite{Billo:2013fi,Billo:2013jba,Seiberg:1994aj,Bajc:2025jjv}, $\tilde\alpha_{2n}$ must be a quasi-modular form of weight $2n$, invariant under the $SO(8)$ flavor symmetry. The natural building blocks are therefore the mass invariants $R,T_1,T_2,N$ together with the Eisenstein series $E_2,E_4,E_6$, and because of the odd powers of $q$ noted above, also with the Jacobi theta functions $\theta_2^4,\theta_4^4$.\footnote{The appearance of Jacobi theta functions is natural: the cross-ratio entering the Heun equation is itself $t=\theta_2^4/\theta_3^4$.} This is, however, not immediate: $R,N,E_2,E_4,E_6$ are already good building blocks, being both $SO(8)$-invariant and (quasi-)modular, but the same is not true of $T_1,T_2$ or of $\theta_2^4,\theta_4^4$ individually, which mix into one another under the modular group. We must therefore first identify which combinations of $T_1,T_2,\theta_2^4,\theta_4^4$ are simultaneously $SO(8)$-invariant and modular covariant, before we can use them to build $\tilde\alpha_{2n}$.

To this end, let us track how the mass invariants and the Jacobi theta functions transform under the modular group. Recall $q=\exp(i\pi\tau_0)$, with $\tau_0$ the complexified gauge coupling, transforming under the generators of the modular group as
\begin{equation}
    \mathcal{T}: \tau_0\rightarrow \tau_0+1,  \quad \mathcal{S}: \tau_0\rightarrow -\frac{1}{\tau_0}.
\end{equation}
The Jacobi theta functions transform as
\begin{equation}
\begin{split}
    \mathcal{T}:&(\theta_2^4(q),\theta_3^4(q),\theta_4^4(q))\rightarrow (-\theta_2^4(q),\theta_4^4(q),\theta_3^4(q)),\\
    \mathcal{S}:&(\theta_2^4(q),\theta_3^4(q),\theta_4^4(q))\rightarrow (-\tau_0^2 \theta_4^4(q),-\tau_0^2\theta_3^4(q),-\tau_0^2\theta_2^4(q)),
    \end{split}
\end{equation} 
and since $t=\theta_2^4(q)/\theta_3^4(q)$, this induces 
\begin{equation}
    \mathcal{T}:t\rightarrow \frac{t}{t-1},\quad \mathcal{S}:t\rightarrow 1-t ,
\end{equation}
with the black brane parameter $t=1/2$ a fixed point of the $\mathcal{S}$ transformation.

On the gauge theory side, these transformations act on the masses (equivalently, on $\vec a$) as \cite{Seiberg:1994aj}:
\begin{equation}
    \mathcal{T}:(m_1,m_2,m_3,m_4)\rightarrow(m_1,m_2,m_3,-m_4) \text{ or } (a_0,a_t,a_1,a_\infty)\rightarrow(a_0,a_t,a_\infty,a_1)
\end{equation}
and
\begin{equation}
    \mathcal{S}:\begin{pmatrix} m_1 \\ m_2\\ m_3\\ m_4 \end{pmatrix} \rightarrow  \frac{1}{2} \begin{pmatrix}
        1&1&1&1\\ 1&1&-1&-1\\ 1&-1&1&-1\\ 1&-1&-1&1
    \end{pmatrix}\begin{pmatrix} m_1 \\ m_2\\ m_3\\ m_4 \end{pmatrix} \text{ or } (a_0,a_t,a_1,a_\infty)\rightarrow(a_1,a_t,a_0,a_\infty)
\end{equation}
so that the $SO(8)$ invariants transform as
\begin{equation}
    \mathcal{T}:(R,T_1,T_2,N)\rightarrow (R,T_1,-T_1-T_2,N), \quad \mathcal{S}:(R,T_1,T_2,N)\rightarrow (R,T_2,T_1,N).
\end{equation}
Note that $R$ and $N$ are invariant, while $T_1,T_2$ mix into each other. Likewise, $\theta_2^4$ and $\theta_4^4$ permute into one another (with an extra weight-2 automorphy factor under $\mathcal{S}$), so neither is separately a modular form. Let us define the two objects
\begin{equation}
    V_T=(T_1, T_2), \quad V_{\theta}=(\theta_2^4+2 \theta_4^4,-2\theta_2^4-\theta_4^4),
\end{equation}
each transforming as a two-dimensional representation of the modular group. Explicitly, a group element $g\in PSL(2,\mathbb{Z})$ acts on them as
\begin{equation}
    g: V_i \rightarrow \chi_i(g)\, \pi(g)\, V_i\, , \qquad i=T,\theta\, ,
\end{equation}
where $\chi_i$ is the automorphy factor: $\chi_T=1$ identically, while $\chi_\theta(\mathcal{T})=1$ , $\chi_\theta(\mathcal{S})=\tau_0^2$, and $\pi(g)$ is a pure permutation matrix, common to both $V_T$ and $V_\theta$:
\begin{equation}
    \pi(\mathcal{S})=\begin{pmatrix}
        0 &1\\1&0
    \end{pmatrix}, \quad  \pi(\mathcal{T})=\begin{pmatrix}
        1 &0\\-1&-1
    \end{pmatrix}\, .
\end{equation}
The action of the modular group thus factorizes into an automorphy factor and a permutation. Since $\pi(\mathcal{S})^2=\mathds{1}$, $\pi(\mathcal{T})^2=\mathds{1}$ and $(\pi(\mathcal{S})\pi(\mathcal{T}))^3=\mathds{1}$, the permutation matrices generate a finite group isomorphic to $S_3$.\footnote{For the mass invariants this is a manifestation of triality: the Dynkin diagram of $SO(8)$ has an outer-automorphism symmetry permuting the three inequivalent 8-dimensional representations (vector $8_V$, spinor $8_S$, conjugate spinor $8_C$), giving the symmetry group $S_3$. For the Jacobi theta functions it instead reflects the fact that $t=\theta_2^4/\theta_3^4$ is the cross-ratio of four points $(t,1;0,\infty)$: permuting these four points generates $S_3$, since swapping two pairs simultaneously leaves the cross-ratio unchanged.}

Because the action factorizes this way, any $S_3$-invariant polynomial built purely from the components of $V_T$ and $V_\theta$ is left invariant by the permutation part $\pi(g)$ by construction, and so under a modular transformation it can only pick up the automorphy factor $\chi_i(g)$, to a power fixed by its degree, which is exactly the covariance property that defines a (quasi-)modular form of that weight. We can therefore obtain proper modular building blocks for $\tilde\alpha_{2n}$ by constructing the $S_3$-invariants of $V_T$ and $V_\theta$.

Since for any finite group $G$ acting linearly on a finite-dimensional vector space $V$ the ring of invariant polynomials $\mathbb{C}[V]^G$ is finitely generated, we can ask how many invariants generate $\mathbb{C}[V_T \oplus V_\theta]^{S_3}$. This is answered by the Molien series (reviewed in Appendix~\ref{app:molien}), whose coefficient of $T^n\theta^m$ counts the number of independent invariants built from $n$ copies of $V_T$ and $m$ copies of $V_\theta$: 
\begin{equation}
    G(T,\theta)=\frac{1}{|S_3|} \sum_{g\in S_3}\frac{1}{\det(1-Tg)\det(1-\theta g)},
\end{equation}
which evaluates to
\begin{equation}
    G(T,\theta)=\frac{1+T\theta+T^2\theta+T\theta^2+T^2\theta^2+T^3\theta^3}{(1-T^2)(1-T^3)(1-\theta^2)(1-\theta^3)}.
\end{equation}

The denominator identifies four \emph{primary} invariants -- algebraically independent invariants, with no relations among themselves -- of degree $2$ and $3$ in $V_T$ alone, and of degree $2$ and $3$ in $V_\theta$ alone. These do not generate every invariant by themselves: any invariant is instead a polynomial in the four primaries times one of finitely many \emph{secondary} invariants, added together. The numerator's six terms tell us how many secondaries there are and their degree: one of degree $0$ (the constant $1$), and one each of bidegree $(1,1),(2,1),(1,2),(2,2),(3,3)$ in $(T,\theta)$.

To construct the invariants, we embed both $T_1,T_2$ and $\theta_2^4,\theta_4^4$ into a three-component vector summing to zero,
\begin{equation}
    \hat V_T=(T_1,T_2,-T_1-T_2)\, , \qquad
    \hat V_\theta=(\theta_2^4+2\theta_4^4,\,-2\theta_2^4-\theta_4^4,\,\theta_2^4-\theta_4^4)\, ,
\end{equation}
on which $S_3$ acts simply by permuting the index $i=1,2,3$, the same way in both vectors. This makes invariants easy to write down: for any $a,b\geq0$,
\begin{equation}
    Q_{a,b}\equiv\sum_{i=1}^3 (\hat V_T)_i^a (\hat V_\theta)_i^b
\end{equation}
is automatically $S_3$-invariant, since permuting the index just permutes the terms being summed.

For any three quantities summing to zero, it is a standard fact that the ring of permutation-invariant polynomials in them is freely generated by just two invariants: their pairwise combination at degree $2$ and their product at degree $3$. Applied to $\hat V_T$, this gives
\begin{equation}
\label{UV}
    \mathbb{C}[V_T]^{S_3}=\mathbb{C}[U,V]\, , \ \ U\equiv\tfrac12 Q_{2,0}=T_1^2+T_2^2+T_1T_2\, , \ \ V\equiv-\tfrac13 Q_{3,0}=T_1T_2(T_1+T_2)\, , 
\end{equation}
while applied to $\hat V_\theta$, and using the classical identities $E_4=\theta_2^8+\theta_4^8+\theta_2^4\theta_4^4$ and $E_6=(\theta_4^4-\theta_2^4)(2\theta_2^4+\theta_4^4)(\theta_2^4+2\theta_4^4)/2$, it gives
\begin{equation}
    \mathbb{C}[V_\theta]^{S_3}=\mathbb{C}[E_4,E_6]\, , \qquad E_4\equiv\tfrac16 Q_{0,2}\, , \qquad E_6\equiv\tfrac16 Q_{0,3}\, .
\end{equation}
Together, $U,V,E_4,E_6$ are the four primary invariants identified by the Molien series.

The same construction, at mixed bidegree, produces the secondary invariants: taking $(a,b)=(1,1),(2,1),(1,2)$ gives the three genuinely new, mixed invariants
\begin{equation}
\begin{split}
M_2&\equiv \tfrac13 Q_{1,1}=T_1 \theta_4^4-T_2 \theta_2^4\, , \\
M_4&\equiv \tfrac13 Q_{1,2}= T_1 \theta_4^8+2(T_1+T_2) \theta_2^4 \theta_4^4 +T_2 \theta_2^8\, ,\\
P_2&\equiv Q_{2,1}=2 T_1 T_2(\theta_2^4-\theta_4^4)+T_1^2(2\theta_2^4+\theta_4^4)-T_2^2(\theta_2^4+2\theta_4^4)\, .
\end{split}
\end{equation}
The two remaining bidegrees required by the Molien series, $(2,2)$ and $(3,3)$, need no new invariant: using
\begin{equation}
    Q_{2,2}=\tfrac16 Q_{2,0}Q_{0,2}+\tfrac13 Q_{1,1}^2\, , \qquad
    Q_{3,3}=\tfrac13 Q_{3,0}Q_{0,3}+\tfrac32 Q_{1,1}Q_{2,2}-\tfrac12 Q_{1,1}^3\, ,
\end{equation}
both $Q_{2,2}$ and $Q_{3,3}$ reduce to polynomials in the primaries $U,V,E_4,E_6$ together with $M_2^2$ and $M_2^3$ respectively. So the six secondary invariants required by the Molien series are the constant $1$, the three genuinely new invariants $M_2,M_4,P_2$, and the two powers $M_2^2,M_2^3$.

By mass dimension and modular weight of $T_1,T_2$ (dimension $4$) and of $\theta_2^4,\theta_4^4$ (dimensionless), $M_2$ is a modular form of weight $2$ and mass dimension $4$, $M_4$ is a modular form of weight $4$ and mass dimension $4$, and $P_2$ is a modular form of weight $2$ and mass dimension $8$.

We now have every ingredient needed to write down the general ansatz. Since $\tilde\alpha_{2n}$ is a quasi-modular form of weight $2n$ and mass dimension $2n+2$, it must take the form
\begin{equation}
\begin{split}
\tilde\alpha_{2n}(q) &=
 \sum_{i=0}^{3}\sum_{i+k+2l+3m=n}
\tilde c^{(i)}_{klm} M_2^i(q) E_2^k(q) E_4^l(q) E_6^m(q) \\
&\quad + \sum_{1+k+2l+3m=n}
\tilde c^{(4)}_{klm} P_2(q) E_2^k(q) E_4^l(q) E_6^m(q) \\
&\quad + \sum_{2+k+2l+3m=n}
\tilde c^{(5)}_{klm} M_4(q) E_2^k(q) E_4^l(q) E_6^m(q),
\label{alphatildeGF}
\end{split}
\end{equation}
Here and below we write $\tilde\alpha_{2n}(q)$, suppressing its dependence on $\nu$ and $\omega$, since in this section we are only interested in how $\tilde\alpha_{2n}$ depends on $q$. The dependence on $\nu$ and $\omega$ enters only through the mass invariants $R,T_1,T_2,N$, both inside $M_2,M_4,P_2$ and in the coefficients $\tilde c^{(i)}_{klm}$.

The above coefficients $\tilde{c}^{(i)}_{klm}$ can be further expanded as ($i=0,\ldots,5$)
\bea
\label{ccoeffs}
\tilde c^{(i)}_{klm} &=& \sum_{0\leq w_i+d\leq n+1} \ \sum_{4u+6v+r+3s=d} c^{(i)}_{klmuvrs} U^u V^v R^r N^s \\
w_{0,1,2,3,4,5}&=&(0,2,4,6,4,2)
\eea
although not all terms in this expansion are independent. In fact, the invariants $U$, $V$, $R$, $N$ \eqref{UV}, \eqref{R}-\eqref{N} are functions of the 
four masses $m_i$ (\ref{mass}) which are parametrised by only $\omega$ and $\nu$.

The coefficients $\tilde c^{(i)}_{klm}$ (and then through them, the more fundamental coefficients $c^{(i)}_{klmuvrs}$) are fixed by matching to the explicit $q$-series generated by Zamolodchikov's recursion. However, as shown in \cite{Billo:2013fi,Billo:2013jba}, \cite{Bajc:2025jjv}, the quasimodular forms $\tilde \alpha_{2n}$ obey a modular anomaly equation,
\beq
\frac{\partial\tilde\alpha_{2l}}{\partial E_2}=\frac{(4R-1)^2}{96}\delta_{l,1}+\left(\frac{l-1}{6}\right)\left(\frac{4R-1}{2}\right)\tilde \alpha_{2(l-1)}
+\frac{1}{6}\sum_{i=1}^{l-2}i(l-i-1)\tilde\alpha_{2i}\tilde\alpha_{2l-2i-2},
\label{anomalyeq}
\eeq
which recursively determines every coefficient multiplying a positive power of $E_2$, i.e.\ $\tilde c^{(i)}_{klm}$ with $k>0$. The only genuinely unknown coefficients are therefore the $E_2$-independent ones, $\tilde c^{(i)}_{0lm}$, which we fix by matching to Zamolodchikov's recursion in the next section.

\section{Zamolodchikov's $q$-recursion \label{sec:zam}}

We now turn to the explicit computation that fixes the remaining, $E_2$-independent coefficients in the modular ansatz \eqref{alphatildeGF}, by matching it against the low-order $q$-expansion
of $H$ obtained from Zamolodchikov’s recursion \cite{Zamolodchikov:1984eqp,Zamolodchikov:1987avt,Zamolodchikov:1995aa,Poghossian:2009mk}, reviewed in Appendix \ref{ap:Zamolodchikov}.

As in \cite{Bajc:2025jjv}, we rewrite the Zamolodchikov $q$-recursion into an explicit form
\begin{equation}
\mathcal{H}(q,\Delta,\vec{\mu})=1+\sum_{\chi=1}^\infty c_\chi(a,\vec{\mu},\epsilon_j) q^{\chi},
\label{explicitZamolodchikov}
\end{equation}
where
\begin{equation}
c_\chi(a,\vec{\mu},\epsilon_j)=\sum_{k=1}^\chi \sum_{\sum_{i=1}^k m_i n_i=\chi} \frac{\epsilon_1 \epsilon_2}{\left(-a^2+\left(\frac{m_1\epsilon_1+n_1 \epsilon_2}{2}\right)^2\right)} \frac{\prod_{i=1}^k R_{m_i n_i}(\vec{\mu})}{\prod_{i=1}^{k-1}( \Delta_{m_i n_i}+m_i n_i -\Delta_{m_{i+1} n_{i+1}})},
\end{equation}
with $\Delta_{m,n}$ and $R_{m,n}$ explicitly stated in Appendix \ref{ap:Zamolodchikov}.
Recalling that
\begin{equation}
H(a,q,\nu)=\lim_{\epsilon_1 \rightarrow 1,\epsilon_2 \rightarrow 0}\epsilon_1 \epsilon_2 \log \mathcal{H}(a,q,\nu)=\sum_{n=1}^\infty \frac{\alpha_{2n}(q,\nu,\omega)}{a^{2n}},
\label{H}
\end{equation}
we obtain the coefficients $\alpha_{2n}(q,\nu,\omega)$ by plugging \eqref{explicitZamolodchikov} into \eqref{H} and expanding for large $a$. Unlike in \cite{Bajc:2025jjv}, this $q$-series no longer truncates to even powers of $q$
only, since $R_{m,n}\neq0$ for $m,n$ both odd if $\omega \neq 0$.

To assemble $\tilde\alpha_{2n}$ from $\alpha_{2n}$ we also need its $q$-independent piece, coming from the perturbative (one-loop) part of the prepotential, which we found to be\footnote{This coefficient could be also found by  using the same procedure as described in this section, just one power of $q$ further than needed for $\alpha_{2n}$ itself, for the first few $n$ and by matching with the $\omega=0$ case found in \cite{Bajc:2025jjv}. As a check, since $q=0$ gives $\mathcal{H}=1$ and hence $\alpha_{2n}=0$, while $E_{2,4,6}(q=0)=1$, $\theta_4(q=0)=1$, $\theta_2(q=0)=0$, the ansatz \eqref{alphatildeGF} correctly reduces at $q=0$ to $f_{2n}=\sum_{i=0}^3 \sum_{i+k+2l+3m=n} \tilde c^{(i)}_{klm} T_1^i+\sum_{1+k+2l+3m=n} \tilde c^{(4)}_{klm} (T_1^2-2 T_1 T_2-2 T_2^2)+\sum_{2+k+2l+3m=n} \tilde c^{(5)}_{klm} T_1$.}
\begin{equation}
    f_{2n}(\nu,\omega)=\frac{B_{2n+2}\big(\frac{1+\nu}{2}-\frac{\omega}{4\pi}\big)+B_{2n+2}\big(\frac{1+\nu}{2}+\frac{\omega}{4\pi}\big)+2 B_{2n+2}\big(\frac{1}{2}+i\frac{\omega}{4\pi}\big)-2^{-2n} B_{2n+2}}{2n(2n+1)(n+1)}\, ,
\end{equation}
Altogether,
\begin{equation}
    \tilde \alpha_{2n}(q,\nu, \omega)=\alpha_{2n}(q,\nu, \omega)+f_{2n}(\nu,\omega)\, .
\end{equation}
The coefficients $\alpha_{2n}(\nu,\omega)$ and $\tilde\alpha_{2n}(\nu,\omega)$ entering the propagator in Section~\ref{sec:prop} are obtained from these by setting the instanton counting parameter to its black brane value,
\begin{equation}
    \alpha_{2n}(\nu,\omega)\equiv\alpha_{2n}(q,\nu,\omega)\big|_{q=e^{-\pi}}\, , \qquad \tilde\alpha_{2n}(\nu,\omega)\equiv\tilde\alpha_{2n}(q,\nu,\omega)\big|_{q=e^{-\pi}}\, .
\end{equation}
Keeping $q$ arbitrary is what allows us to use modularity to resum all instanton contributions, and only at the end do we set $q=e^{-\pi}$.

We can now close the loop with Section~\ref{sec:modular}. In the ansatz \eqref{alphatildeGF}, all coefficients multiplying a positive power of $E_2$, i.e.\ $\tilde c^{(i)}_{klm}$ with $k>0$, are already fixed recursively by the anomaly equation \eqref{anomalyeq} and only the $E_2$-independent, purely modular coefficients remain unknown. Splitting accordingly
\begin{equation}
    \tilde \alpha_{2n}(q)=\tilde \alpha_{2n}^{E_2}+\tilde \alpha_{2n}^{\text{modular}}\, ,
    \label{alphatildesplit}
\end{equation}
where, as in \eqref{alphatildeGF}, we only display the dependence on $q$, and the anomaly equation determines $\tilde\alpha_{2n}^{E_2}$, while \eqref{alphatildeGF} restricted to $E_2^0$ gives the general form of $\tilde\alpha_{2n}^{\text{modular}}$ in terms of the finitely many unknown coefficients $\tilde c^{(i)}_{0lm}$. To fix them, we use
that the left-hand side of \eqref{alphatildesplit} is known explicitly as a $q$-series, computed from \eqref{explicitZamolodchikov}--\eqref{H} together with $f_{2n}$ and expanding both the Eisenstein series and the Jacobi theta functions on the right-hand side as $q$-series as well turns \eqref{alphatildeGF} into a $q$-series with the finitely many unknown coefficients $\tilde c^{(i)}_{0lm}$ as its only free parameters (as $\tilde c^{(i)}_{klm}$ with $k>0$ are already known from the anomaly equation). Matching the two $q$-series term by term, up to a sufficiently high order in $q$,\footnote{Modular weight allows $C_0+2C_1+2C_2+C_3=n+1$ coefficients $\tilde c^{(i)}_{0lm}$, where $C_i=\big|\{m:\ 0\leq m\leq\lfloor\frac{n-i}{3}\rfloor,\ n-i-m\equiv 0 \bmod 2\}\big|$. For $n\leq3$ one of them is forbidden by mass dimension, namely the coefficient of $P_2$, $M_2^2$ and $M_2^3$ for $n=1,2,3$ respectively. The number of unknowns is therefore $N=n$ for $n\leq3$ and $N=n+1$ for $n>3$, and we need the $q$-expansion up to order $q^N$.} then completely determines all the coefficients $\tilde c^{(i)}_{0lm}$ appearing in \eqref{alphatildeGF}, and hence $\tilde\alpha_{2n}(q,\nu,\omega)$ itself.

The first few $\tilde\alpha_{2n}(q,\nu,\omega)$ are
\begin{align}
   & \tilde \alpha_{2}=-M_2 + \frac{(1 - 4 R)^2}{96} E_2\\
    & \tilde \alpha_{4}=\frac{(1 - 4 R)}{12} E_2 M_2 + \frac{(1 - R)}{6} M_4 + \frac{(4 R-1)^3}{2304}E_2^2 +\left( \frac{N}{5} + \frac{(1 - 4 R)^2 (4 R-13)}{11520} \right)E_4\\
    & \tilde \alpha_{6}=\frac{1}{6}E_2 M_2^2 + \frac{1}{9}E_4 P_2 - \frac{5(1 - 4R)^2}{576}E_2^2 M_2 + \frac{5(1 - 4R)^4}{165888}E_2^3 + \frac{(5 - 4R)R-1}{36}E_2 M_4 \notag \\ & 
    -\frac{(4R-7)(4R-3)}{192}E_4 M_2 + \frac{(4R-1)\big(2304N + (1-4R)^2(4R-13)\big)}{69120}E_2 E_4\\
    &+\left(\frac{2}{105}N(3R-7) + \frac{(1-4R)^2\big(3323 + 8R(22R-227)\big)}{5806080} + \frac{8U}{63}\right)E_6 \notag
\end{align}
and two more can be found in Appendix \ref{app:alphatilde}.

\section{Final Result and Conclusions \label{sec:conclusions}}
After we obtained $\tilde \alpha_{2n}$, we can follow the procedure described in section \ref{sec:prop} to obtain the low temperature expansion of the propagator dual to the AdS black brane:
\begin{equation}
\begin{split}
    &G_R/G_R^0=1 - \frac{2\pi^4}{15} \frac{\Gamma(\nu+3)}{\Gamma(\nu-2)}\left(\frac{4\omega^2}{(\vec{k}^2-\omega^2)^3} +\frac{1}{(\vec{k}^2-\omega^2)^2}\right)T^4 \\
&+ \frac{2\pi^8}{1575}\Gamma(\nu+3)\left(\frac{16\big(44+\nu(34+7\nu)\big)\omega^4}{\Gamma(\nu-5)(\vec{k}^2-\omega^2)^6}+ \frac{8\big(-340+\nu(-196+\nu(-11+7\nu))\big)\omega^2}{\Gamma(\nu-4)(\vec{k}^2-\omega^2)^5}\right.\\
&\left.+ \frac{\big(1000+\nu(324+\nu(-62+\nu(-9+7\nu)))\big)}{\Gamma(\nu-3)(\vec{k}^2-\omega^2)^4}\right)T^8\\
&-\frac{4\pi^{12}}{3378375}\Gamma(\nu+3)\left(\frac{64\big(91880+\nu(111672+\nu(52784+143\nu(81+7\nu)))\big)\omega^6}{3 \Gamma(\nu-8)(\vec{k}^2-\omega^2)^9}\right.\\
&+\frac{16\big(-1206880+\nu(-1316496+\nu(-524100+\nu(-79880+143\nu(5+7\nu))))\big)\omega^4}{ \Gamma(\nu-7)(\vec{k}^2-\omega^2)^8}\\
&+\frac{4\big(9873520+\nu(9362112+\nu(2902344+\nu(182040+\nu(-55905+143\nu(-24+7\nu)))))\big)\omega^2}{\Gamma(\nu-6)(\vec{k}^2-\omega^2)^7}\\
&\left.+\frac{\big(4561360+\nu(4318992+\nu(1416624+\nu(166500+\nu(8535+143\nu(36+7\nu)))))\big)}{3\Gamma(\nu-6)(\vec{k}^2-\omega^2)^6}\right)T^{12}\\
&+\mathcal{O}(T^{16}).
\end{split}
\end{equation}
Only powers $T^{4n}$ appear, as expected from the stress-tensor sector of the thermal OPE in $d=4$. This is a nontrivial property of our computation: odd powers of $q$ enter the individual $\tilde\alpha_{2n}$, and $\omega$ reshuffles terms between orders in $1/a$ and orders in $T$, yet all other powers of $T$ cancel. At each order the $\omega$-dependence enters only through $\omega^2/(\vec k^2-\omega^2)$, with at most $n$ powers of it at order $T^{4n}$. The coefficients are polynomials in $\nu$, and at $\omega=0$ the result reduces to that of \cite{Bajc:2025jjv}.
 
The result agrees, through order $T^{12}$, with the low-temperature expansion obtained from the near-boundary expansion of the bulk wave equation \cite{Fitzpatrick:2019zqz}. This is a nontrivial cross-check: that computation never solves the connection problem between the boundary and the horizon, whereas we obtained the result by mapping the connection problem onto the Coulomb branch of $\mathcal{N}=2$ $SU(2)$ SYM with $N_f=4$ and exploiting its electric-magnetic duality. The agreement therefore tests not only the numerical coefficients but the identification itself: the finite-$\omega$ black brane propagator is governed by the $S$-duality of this gauge theory, now with all four hypermultiplet masses switched on.
 
Conceptually, the following changed relative to $\omega=0$ \cite{Bajc:2025jjv}. There, only two hypermultiplet masses are non-zero, only even powers of $q$ appear in Zamolodchikov's recursion, and the whole expansion can be written with $E_2,E_4,E_6$ alone. Turning on $\omega$ activates all four masses and the full $SO(8)$ flavour symmetry. Odd powers of $q$ appear, and with them the Jacobi theta functions $\theta_2^4,\theta_4^4$. Neither these nor the quartic mass invariants $T_1,T_2$ are separately modular covariant. The key observation is that the two pairs transform in the same way. Each parameter $a_i$ is the local exponent attached to one of the singular points $0,t,1,\infty$ of the Heun equation. A modular transformation acts on the cross-ratio $t=\theta_2^4/\theta_3^4$ as an anharmonic transformation, i.e.\ a permutation of the singular points, and therefore simultaneously permutes the $a_i$. On the masses, this induced $S_3$ is precisely the triality outer automorphism of $SO(8)$. As a result, the modular group acts on $(T_1,T_2)$ and on $(\theta_2^4+2\theta_4^4,\,-2\theta_2^4-\theta_4^4)$ through the same permutation matrices, up to an automorphy factor. Every polynomial in the mass invariants and theta functions that is invariant under the joint $S_3$ is therefore automatically a (quasi-)modular form. The Molien series counts these invariants. The invariant ring is generated by the four primary invariants $U,V,E_4,E_6$ together with the secondary invariants $1,M_2,M_4,P_2,M_2^2,M_2^3$, of which $M_2,M_4,P_2$ are genuinely new mixed invariants built from both masses and theta functions. Together with $R$, $N$ and $E_2$, they fix the most general ansatz \eqref{alphatildeGF} for $\tilde\alpha_{2n}$. The modular anomaly equation \eqref{anomalyeq} then determines recursively all coefficients multiplying a positive power of $E_2$. What remains at each order is a finite number of purely modular, $E_2$-independent coefficients, fixed by matching the ansatz against a finite number of terms in Zamolodchikov's $q$-recursion. In this way the full instanton sum at a given order in $T$ is obtained from a finite computation, even though at the black brane point $q=e^{-\pi}$, the fixed point of $\mathcal{S}$, no instanton contribution is parametrically suppressed.
 
Several natural extensions remain. First, we worked at fixed $\omega/|\vec k|<1$; the opposite regime, including the approach to the light cone, deserves a separate study. Second, the same strategy should apply to other black hole backgrounds whose perturbation equations reduce to Heun equations, such as charged or rotating black holes, or black branes in other dimensions. There the cross-ratio is generally not at the self-dual point $t=1/2$, so the full modular group, and not only its $\mathcal{S}$-fixed point, would be probed. Third, correlators of other operators on the same black brane, such as conserved currents and the stress tensor, are governed by Heun equations with different local exponents. In the gauge theory these correspond to different values of the four hypermultiplet masses, while the triality structure and the modular ansatz constructed here carry over unchanged, so their low-temperature expansions should be accessible with the same methods.

\subsubsection*{Acknowledgments}
We thank Adrian Lugo for reading the manuscript and comments. BB and KT acknowledge the financial support from the Slovenian 
Research Agency (research core funding No.~P1-0035).

\appendix

\section{Eisenstein series and Jacobi theta functions\label{app:EisensteinJacobi}}
In this Appendix we collect the definitions, transformation properties, and
special values of the Jacobi theta functions and Eisenstein series used
throughout this paper, with $q=\exp(i\pi\tau)$.
\subsection*{Jacobi theta functions}

The Jacobi theta functions are defined by
\begin{equation}
\theta_2(q) = \sum_{n=-\infty}^{\infty} q^{(n+1/2)^2}\,,
\qquad
\theta_3(q) = \sum_{n=-\infty}^{\infty} q^{n^2}\,,
\qquad
\theta_4(q) = \sum_{n=-\infty}^{\infty} (-1)^n q^{n^2}\,,
\end{equation}
which satisfy the identity
\begin{equation}
\theta_3^4(q) = \theta_2^4(q) + \theta_4^4(q)\,.
\end{equation}
Their logarithmic $q$-derivatives can be written in terms of the Eisenstein
series introduced below,
\begin{equation}
q\partial_q \log\theta_2^4 = \frac{1}{3}\left(E_2+\theta_3^4+\theta_4^4\right),
q\partial_q \log\theta_3^4 = \frac{1}{3}\left(E_2+\theta_2^4-\theta_4^4\right),
q\partial_q \log\theta_4^4 = \frac{1}{3}\left(E_2-\theta_2^4-\theta_3^4\right).
\label{thetader}
\end{equation}

\subsection*{Eisenstein series}

The Eisenstein series of weight $2k$ are defined as
\begin{equation}
E_{2k}(\tau) = \frac{1}{2\zeta(2k)}\sum_{(m,n)\in\mathbb{Z}^2\setminus\{(0,0)\}} \frac{1}{(m+n\tau)^{2k}}\,,
\end{equation}
which for $k>1$ converges absolutely to a holomorphic function of $\tau$ on the upper
half-plane $\mathbb{H}$. For $k>1$, $E_{2k}(\tau)$ is a modular form of
weight $2k$, i.e.\ it is $PSL(2,\mathbb{Z})$-covariant,
\begin{equation}
\tau \to \frac{a\tau+b}{c\tau+d}\,,
\qquad
E_{2k}(\tau) \to (c\tau+d)^{2k}\,E_{2k}(\tau)\,,
\qquad a,b,c,d\in\mathbb{Z}\,,\ ad-bc=1\,.
\end{equation}
The series $E_4$ and $E_6$ generate the ring of modular forms.
The weight-2 series $E_2$ is not itself a modular form; under
$PSL(2,\mathbb{Z})$ it transforms with an inhomogeneous (anomalous) term,
\begin{equation}
E_2(\tau) \to (c\tau+d)^2\,E_2(\tau) - \frac{6i}{\pi}\,c(c\tau+d)\,,
\end{equation}
and is therefore only \emph{quasi}-modular, of weight 2. The ring of
quasi-modular forms is $\mathbb{C}[E_2,E_4,E_6]$.

The Fourier ($q$-)expansions of the first three Eisenstein series are
\begin{equation}
E_2(\tau) = 1-24\sum_{n=1}^{\infty}\sigma_1(n)\,q^{2n} = 1-24\sum_{n=1}^{\infty}\frac{n\,q^{2n}}{1-q^{2n}}\,,
\end{equation}
\begin{equation}
E_4(\tau) = 1+240\sum_{n=1}^{\infty}\sigma_3(n)\,q^{2n} = 1+240\sum_{n=1}^{\infty}\frac{n^3\,q^{2n}}{1-q^{2n}}\,,
\end{equation}
\begin{equation}
E_6(\tau) = 1-504\sum_{n=1}^{\infty}\sigma_5(n)\,q^{2n} = 1-504\sum_{n=1}^{\infty}\frac{n^5\,q^{2n}}{1-q^{2n}}\,,
\end{equation}
where $\sigma_k(n)=\sum_{d|n} d^k$ is the divisor function.

The ring of quasi-modular forms is closed under the operator
$q\partial_q$, hence the derivatives of $E_2,E_4,E_6$ can again be expressed within
$\mathbb{C}[E_2,E_4,E_6]$:
\begin{equation}
q\,\partial_q E_2 = \frac{E_2^2-E_4}{6}\,,
\quad
q\,\partial_q E_4 = \frac{2}{3}\left(E_2E_4-E_6\right)\,,
\quad
q\,\partial_q E_6 = E_2E_6-E_4^2\,.
\label{eisensteinder}
\end{equation}
\subsection*{Relation between Eisenstein series and theta functions}

The weight-4 and weight-6 Eisenstein series can be written directly in terms
of the Jacobi theta functions,
\begin{equation}
\begin{split}
E_4(q) &= \frac{1}{2}\left(\theta_2(q)^8+\theta_3(q)^8+\theta_4(q)^8\right)\,,
\\
E_6(q) &= \frac{1}{2}\sqrt{\frac{1}{2}\left(\theta_2(q)^8+\theta_3(q)^8+\theta_4(q)^8\right)^3 - 27\left(\theta_2(q)\theta_3(q)\theta_4(q)\right)^8}\,,
\end{split}
\end{equation}
which in turn can be expressed through the Dedekind eta function,
\begin{equation}
\theta_2(e^{i\tau\pi}) = \frac{2\eta^2(2\tau)}{\eta(\tau)}\,,
\qquad
\theta_3(e^{i\tau\pi}) = \frac{\eta^5(\tau)}{\eta^2\!\left(\frac{1}{2}\tau\right)\eta^2(2\tau)}\,,
\qquad
\theta_4(e^{i\tau\pi}) = \frac{\eta^2\!\left(\frac{1}{2}\tau\right)}{\eta(\tau)}\,.
\label{thetaeta}
\end{equation}

\subsection*{Special values at $\tau=i$}

The point $q=\exp(-\pi)$, corresponding to $\tau=i$, will play a distinguished
role in the main text as it corresponds to $t=1/2$ which is the value for the black brane. The Dedekind eta function takes the special values
\begin{equation}
\eta(i)=\frac{\Gamma\!\left(\frac{1}{4}\right)}{2\pi^{3/4}}\,,
\qquad
\eta\!\left(\tfrac{1}{2}i\right)=\frac{\Gamma\!\left(\frac{1}{4}\right)}{2^{7/8}\pi^{3/4}}\,,
\qquad
\eta(2i)=\frac{\Gamma\!\left(\frac{1}{4}\right)}{2^{11/8}\pi^{3/4}}\,,
\end{equation}
from which, using \eqref{thetaeta}, one finds for the theta functions
\begin{equation}
\theta_2^4(e^{-\pi}) = \theta_4^4(e^{-\pi}) = \frac{\pi}{2\,\Gamma(3/4)^4}\,,
\qquad
\theta_3^4(e^{-\pi}) = \frac{\pi}{\Gamma(3/4)^4}\,.
\end{equation}
At the same point, the Eisenstein series evaluate to
\begin{equation}
E_2(e^{-\pi}) = \frac{3}{\pi}\,,
\qquad
E_4(e^{-\pi}) = \frac{3\pi^2}{4\,\Gamma(3/4)^8}\,,
\qquad
E_6(e^{-\pi}) = 0\,.
\label{eq:eisenstein-special}
\end{equation}

We note that, although $E_6(e^{-\pi})=0$, its derivative $q\partial_q E_6$
does not vanish at this point and must be kept in any computation
involving derivatives evaluated at $q=e^{-\pi}$.

\section{Zamolodchikov's $q$-recursion \label{ap:Zamolodchikov}}
The Virasoro conformal block admits no closed-form expression, but it can be
generated order by order in a convenient expansion parameter through a
recursion relation due to Zamolodchikov \cite{Zamolodchikov:1984eqp,Zamolodchikov:1987avt,Zamolodchikov:1995aa,Poghossian:2009mk}. Here we collect the ingredients
needed to state it, following the AGT dictionary that connects the block to
the instanton expansion of an $\mathcal{N}=2$ gauge theory partition
function.

The 4-point block of a $2d$ CFT, with external primaries of dimension
$\Delta_{1,2,3,4}$ inserted at $0,t,1,\infty$ and an internal primary of
dimension $\Delta$ exchanged in the channel, is conventionally written as
\begin{equation}
\begin{split}
\mathcal{F}(\Delta,\Delta_i,t) =& (16q)^{-\alpha^2}\, t^{\,Q^2/4-\Delta_1-\Delta_2}(1-t)^{\,Q^2/4-\Delta_1-\Delta_3}\\
&\times \theta_3(q)^{3Q^2-4(\Delta_1+\Delta_2+\Delta_3+\Delta_4)}\, \mathcal{H}(q,\Delta,\vec\mu)\,,
\end{split}
\end{equation}
with $Q$ the background charge, related to the central charge through
$c=1+6Q^2$. It is convenient to trade the conformal dimensions for
momentum-like variables,
\begin{equation}
\Delta_i = \frac{Q^2}{4}-\lambda_i^2\,,\qquad \Delta = \frac{Q^2}{4}-\alpha^2\,,
\end{equation}
with $\alpha$ the internal momentum entering $\mathcal{F}$, and to package
the external data into the four mass-like parameters that appear in
$\mathcal{H}$,
\begin{equation}
\mu_1=\lambda_1+\lambda_2+\frac{Q}{2}\,,\quad
\mu_2=\lambda_1-\lambda_2+\frac{Q}{2}\,,\quad
\mu_3=\lambda_3+\lambda_4+\frac{Q}{2}\,,\quad
\mu_4=\lambda_3-\lambda_4+\frac{Q}{2}\,.
\end{equation}
Under the AGT correspondence, $Q$ is fixed by the two $\Omega$-background
parameters $\epsilon_{1,2}$ via $Q=(\epsilon_1+\epsilon_2)/\sqrt{\epsilon_1\epsilon_2}$,
and the cross-ratio $t$ is traded for the renormalized instanton-counting
parameter $q$ through the elliptic-integral relation
\begin{equation}
q(t)=\exp\!\left(-\pi\,\frac{K(1-t)}{K(t)}\right)\,,\qquad K(t)=\frac{1}{2}\int_0^1 \frac{dx}{\sqrt{x(1-x)(1-tx)}}\,,
\end{equation}
whose inverse can be written compactly in terms of Jacobi theta functions,
\begin{equation}
t(q)=\left(\frac{\theta_2(q)}{\theta_3(q)}\right)^4.
\end{equation}

With these variables in place, the object $\mathcal{H}(q,\Delta,\vec\mu)$
satisfies Zamolodchikov's recursion relation,
\begin{equation}
\mathcal{H}(q,\Delta,\vec\mu) = 1+\sum_{m,n=1}^{\infty} q^{mn}\,\frac{R_{m,n}(\vec\mu)}{\Delta-\Delta_{m,n}}\,\mathcal{H}(q,\Delta_{m,n}+mn,\vec\mu)\,,
\label{Zamolodchikovqrec}
\end{equation}
which expresses $\mathcal{H}$ recursively in terms of
$\mathcal{H}$ evaluated at the shifted dimensions $\Delta_{m,n}+mn$, with
residues $R_{m,n}$ fixed by the degenerate representation content of the
Virasoro algebra. This recursive structure is what makes the $q$-expansion
of the block computable to arbitrarily high order.

The poles $\Delta_{m,n}$ correspond to the dimensions of degenerate
Virasoro representations at level $mn$, and are given in terms of the
momentum-like variables by
\begin{equation}
\Delta_{m,n} = \lambda_{1,1}^2 - \lambda_{m,n}^2\,,
\qquad
\lambda_{m,n} = \frac{m\,\epsilon_1 + n\,\epsilon_2}{2\sqrt{\epsilon_1\epsilon_2}}\,.
\end{equation}
The residues $R_{m,n}(\vec\mu)$ are fixed entirely by the external data
through 
\begin{equation}
R_{m,n}=\frac{2\prod_{r,s} \prod_{i=1}^4 (m_i/\sqrt{\epsilon_1\epsilon_2}-\lambda_{r,s})}{\prod_{k,l}' \lambda_{k,l}},
\end{equation}
where 
$r\in \{ -m+2j-1\ | \ j=1,\ldots, m\}$, $s\in \{ -n+2j-1\ | \ j=1,\ldots, n\}$ and the prime means $(k,l)\in \{ (-m+i, -n+j)\ | \ i=1,\ldots, 2m, \ j=1,\ldots,2n\}\backslash \{ (0,0), (m,n)\} $.

Finally, on the gauge-theory side of the AGT dictionary the internal
momentum $\alpha$ is identified with the Coulomb branch vev $a$ of the
adjoint scalar in the vector multiplet, and the external momenta
$\lambda_i$ with the hypermultiplet mass parameters $\vec a$, according to
\begin{equation}
a = \alpha\sqrt{\epsilon_1\epsilon_2}\,,
\quad
 (a_t,a_0,a_1,a_\infty)=(\lambda_1,\lambda_2,\lambda_3,\lambda_4)\sqrt{\epsilon_1\epsilon_2}\,.
\end{equation}

\section{Molien series \label{app:molien}}
In this appendix we review basic concepts about Molien series \cite{Stanley:1979}.

For a finite group $G$ acting linearly on a vector space $V\cong\mathbb{C}^n$ via $\rho:G\to GL(V)$, the induced action on $\mathbb{C}[V]$ preserves the polynomial degree, so the invariant subring inherits a grading,
\begin{equation}
    \mathbb{C}[V]^G=\bigoplus_{d\geq0}R_d,
\end{equation}
where $R_d$ denotes the vector space of $G$-invariant homogeneous polynomials of degree $d$.
The generating function for the dimensions of this grading, the \emph{Molien series} $M(t)=\sum_{d}\dim(R_d)\,t^d$, has the closed form
\begin{equation}
    M(t)=\frac{1}{|G|}\sum_{g\in G}\frac{1}{\det\big(1-t\,\rho(g)\big)},
    \label{eq:molienformula}
\end{equation}
so the number of invariants at every degree is fixed purely by the character of $\rho$, without constructing any invariant explicitly.
 
Since $G$ is finite and $|G|$ is invertible in $\mathbb{C}$, the Hochster--Eagon theorem guarantees $\mathbb{C}[V]^G$ is Cohen--Macaulay: there exist $n=\dim V$ algebraically independent homogeneous \emph{primary} invariants $\theta_1,\dots,\theta_n$ of degrees $d_1,\dots,d_n$, and finitely many homogeneous \emph{secondary} invariants $\eta_1=1,\dots,\eta_r$ of degrees $e_1=0, e_2,\dots,e_r$, such that every invariant is a unique $\mathbb{C}[\theta_1,\dots,\theta_n]$-linear combination of the $\eta_k$,
\begin{equation}
    \mathbb{C}[V]^G=\bigoplus_{k=1}^r\eta_k\,\mathbb{C}[\theta_1,\dots,\theta_n].
    \label{eq:freemod}
\end{equation}
Counting the basis monomials $\eta_k\theta_1^{a_1}\cdots\theta_n^{a_n}$ of \eqref{eq:freemod} by degree and summing the resulting geometric series in each $a_i$ gives the general rational-function shape
\begin{equation}
    M(t)=\frac{\sum_{k=1}^r t^{e_k}}{\prod_{i=1}^n\left(1-t^{d_i}\right)}.
    \label{eq:molienshape}
\end{equation}
Matching an explicit evaluation of \eqref{eq:molienformula} against \eqref{eq:molienshape} thus suggests the degrees of a possible choice of a full set of primary and secondary invariants directly: the $n$ denominator factors give the $d_i$, and the $r$ numerator terms give the $e_k$. This choice is confirmed by constructing invariants of these degrees and checking that the primaries have no common zero other than the origin, and that the secondaries are linearly independent modulo the ideal of $\mathbb{C}[V]^G$ generated by the primaries.
 
If $G$ is generated by reflections, the Shephard--Todd/Chevalley theorem guarantees that the primaries can be chosen with $r=1$: $\mathbb{C}[V]^G=\mathbb{C}[\theta_1,\dots,\theta_n]$ is itself a polynomial ring, and
\begin{equation}
    M(t)=\frac{1}{\prod_{i=1}^n(1-t^{d_i})}, \qquad \prod_{i=1}^n d_i=|G|.
    \label{eq:reflectionmolien}
\end{equation}
By contrast, if $r>1$, the $n+r-1$ generators $\theta_1,\dots,\theta_n,\eta_2,\dots,\eta_r$ cannot be algebraically independent, since $\mathbb{C}[V]^G$ has Krull dimension $n$, so relations among them are unavoidable.

\section{Higher orders in $\tilde \alpha_{2n}$ \label{app:alphatilde}}

\begin{equation}
\begin{split}
   & \tilde \alpha_{8}=\left(\frac{R}{9}-\frac{35}{144}\right)E_4 M_2^2 + \frac{3R-7}{54}E_6 P_2 + \frac{1267 - 1536N - 4R(693-420R+64R^2)}{17280}E_6 M_2\\
    &+\left(-\frac{N}{9}-\frac{(R-1)(95+8R(4R-17))}{1728}\right)E_4 M_4+\left(\frac{N\big(1673 + 40 R(8R-37)\big)}{10080}\right.\\
    &\left. +\frac{(1-4R)^2\big(4R(14649+4R(76R-1017))-62035\big)}{92897280}+\frac{2}{189}(9R-11)U\right)E_4^2\\
    &+\frac{16R-7}{108}E_2 E_4 P_2 + \left(-\frac{2N}{15}-\frac{(4R-1)(971+8R(94R-239))}{34560}\right)E_2 E_4 M_2\\
    &+\left(\frac{N(3R-7)(4R-1)}{210}+\frac{(4R-1)^3(3323+8R(22R-227))}{23224320}+\frac{2(19R-10)U}{189}\right)E_2 E_6\\
    &-\frac{7(1-4R)^2(R-1)}{1728}E_2^2 M_4 + \frac{7(4R-1)}{144}E_2^2 M_2^2+ \frac{7(4R-1)^5}{2654208}E_2^4 \\
    &-\frac{7(4R-1)^3}{6912}E_2^3 M_2 + \frac{7\big(2304N(1-4R)^2+(1-4R)^4(4R-13)\big)}{3317760}E_2^2 E_4
\end{split}
\end{equation}

\begin{equation}
\begin{split}
   & \tilde\alpha_{10}=\frac{-1361+256N(17-5R)+R\Big(3373-4R\big(675+4R(4R-47)\big)\Big)-1536U}{17280}E_6 M_4\\
    &-\frac{(R-1)(4R-1)^3}{1728}E_2^3 M_4+\left(\frac{11}{64}+\frac{23}{216}R(2R-5)\right)E_2 E_4 M_2^2+\frac{9+7R(2R-5)}{162}E_2 E_6 P_2\\
    &+\frac{(4R-1)^3\big(2304N+(1-4R)^2(4R-13)\big)}{3317760}E_2^3 E_4+\frac{11+16R(5R-4)}{576}E_2^2 E_4 P_2\\
     &-\frac{(4R-1)\Big(9216N+(4R-1)\big(997+8R(98R-253)\big)\Big)}{184320}E_2^2 E_4 M_2+\frac{7(1-4R)^6}{26542080}E_2^5 \\
    &+\frac{3717+8R(58R-325)}{8640}E_6 M_2^2+\frac{8649+16R(103R-460)}{25920}E_4^2 P_2+\left( \frac{20971R}{48384} \right. \\ & \left.-\frac{218297}{1290240}  - \frac{86U}{315}-\frac{32N(596R-1715)+R^2\big(44499+8R(191R-1785)\big)}{120960} \right)E_4^2 M_2\\
    &+\left(\frac{R\Big(18885-2R\big(15603+8R(153R-1057)\big)\Big)}{120960}-\frac{48409}{1935360} - \frac{2}{945}N(83R-77)\right. \\ & \left.  - \frac{8U}{63}\right)E_2 E_6 M_2-\frac{1}{12}E_2^2 M_2^3+\frac{1}{12}E_4 M_2^3+\frac{(1-4R)^2}{96}E_2^3 M_2^2-\frac{7(1-4R)^4}{55296}E_2^4 M_2\\
    &+\frac{128N(11-26R)+R\big(2077-4R(939+4R(28R-153))\big)-3(107+640U)}{17280}E_2 E_4 M_4\\
    &+\left(\frac{13N^2}{525}+\frac{(1-4R)^4\big(3323+8R(22R-227)\big)}{123863040}-\frac{N\big(21+R(8R(51-70R)-177)\big)}{3360}\right. \\ & \left.+\frac{(17-94R)U}{504}\right)E_2^2 E_6
    +\left(\frac{(4R-1)^3\big(4R(12435+4R(68R-879))-51893\big)}{232243200}\right. \\ & \left.+\frac{N\big(4R(20619+4R(2548R-4005))-16821\big)}{302400}+\frac{212 N^2}{4725} +\frac{2(18-67R)U}{567}+\frac{V}{3}\right)E_2 E_4^2\\
    &+\left(\frac{2203N^2}{37125}-\frac{N\big(14037+2R(16R(180-109R)-7911)\big)}{43200}+\frac{(1197-1510R)U}{3240}+\frac{16V}{33}\right.\\ & \left. +\frac{(1-4R)^2\big(3734773+16R(2R(50727+8R(37R-889))-266977)\big)}{2919628800}\right)E_4 E_6
\end{split}
\end{equation}

\printbibliography

\end{document}